\documentclass{pasj00}
\begin{document}
\SetRunningHead{K. Sadakane et al.}{Abundances in the UMi Dwarf Galaxy}
\Received{2004 April 19}
\Accepted{2004 October 27}

\title{Subaru/HDS Abundances in Three Giant Stars in \\
      the Ursa Minor Dwarf Spheroidal Galaxy
\thanks{Based on data collected at Subaru Telescope,
                  which is operated by the National Astronomical Observatory
of Japan.}}

\author{Kozo \textsc{Sadakane}}
\affil{Astronomical Institute, Osaka Kyoiku University, Kashiwara-shi, Osaka
         582-8582}
\email{sadakane@cc.osaka-kyoiku.ac.jp}

\author{Nobuo \textsc{Arimoto}, Chisato \textsc{Ikuta}, Wako \textsc{Aoki}}
\affil{National Astronomical Observatory, 2-21-1 Osawa,
         Mitaka, Tokyo 181-8588}

\author{Pascale \textsc{Jablonka}}
\affil{Observatoire de Paris, 5 place Jules Janssen, F-92195 Meudon Cedex,
France}

\and
\author{Akito \textsc{Tajitsu}}
\affil{Subaru Telescope, National Astronomical Observatory of Japan,
           650 North A'ohoku Place, \\
           Hilo, HI96720, USA}


\maketitle

\begin{abstract}
With the HDS (High Dispersion Spectrograph) on the Subaru telescope, we
obtained high resolution optical region spectra of three red giant stars
(cos 4, cos 82, and cos 347) in the Ursa Minor dwarf spheriodal galaxy.
Chemical abundances  in these stars  have been analysed for 26
elements including $\alpha$--, iron--peak, and neutron capture elements.
All three stars show low abundances of $\alpha$--elements (Mg, Si, and Ca)
and two stars (cos 82 and cos 347) show high abundance of Mn
compared to Galactic halo stars of similar metallicity. One star
(cos 4)  has been confirmed to be very metal deficient ([Fe/H]$=-2.7$) and
found to show anomalously low abundances of Mn, Cu, and Ba. In another star
cos 82 ([Fe/H]$=-1.5$), we have
found large excess of heavy neutron-capture elements with the general
abundance pattern similar to the scaled solar system {\it r}--process 
abundance curve.
These observational results are rather puzzling: low abundances of $\alpha$ 
-- elements
and high abundance of Mn seem to sugggest a significant contribution of SNe Ia
at low metallicity, while there is no hint of {\it s}--process (i.e., AGB
stars) contribution even at [Fe/H]$=-1.5$,
suggesting a peculiar nucleosynthetic history of the UMi dSph galaxy.

\end{abstract}

\section {Introduction}

The origin of the Galactic dwarf spheroidal (dSph) galaxies  is closely
related to the formation and evolutionary history of the Milky Way.
Modern cosmological models based on the Cold Dark Matter paradigm
demonstrate the importance of hierarchical structure formation on all
scales.  Galaxies like the Milky Way and M31 form as part of a local
overdensity in the primordial matter distribution via the agglomeration of
numerous smaller building blocks which independently can develop into dwarf
galaxies.  In the Local Group the leftovers of this process are seen in
the distribution and properties of the dwarf galaxies, with the dwarf
spheroidals found mainly close in to the giant spirals, while the dwarf
irregulars are more evenly distributed throughout the Local Group.

The gravitationally bound dwarf galaxies that have managed to avoid tidal
destruction and subsequent merging have undergone episodic star formation
over a Hubble time.  The relatively gas-rich dwarf irregulars still exhibit
ongoing star formation, while the dwarf spheroidals, being devoid of
significant amounts of gas and dust, are now quiescent and are therefore,
in principle, much simpler systems to study.  The proximity of the
Galactic dSphs offers a unique opportunity for investigating galaxy
formation and evolution in unprecedented detail by studying the photometric
and spectroscopic properties of the dSph stellar populations.

In this respect, an important approach is to explore the chemical
abundances in individual stars belonging to nearby dSph galaxies and to compare
them with abundances found in Galactic halo stars.
Recently, various nearby dSph galaxies have been the subject of extensive
studies concerning accurate abundance analyses based on high dispersion
spectrosopic observations  using ground-based 8-m class telescopes.

\citet{shet1998} analysed high resolution spectra of four red giant stars
in the
Draco dSph galaxy observed with the KECK HIRES spectrograph.
\citet{smc1999} reported preliminay abundances in 14 stars in the
Sagittarius dSph galaxy.
\citet{boni2000} obtained high resolution data of two giant stars in the
Sgr dSph galaxy using the UVES spectrograph on the ESO 8.2 m Kueyen
telescope (VLT).
\citet{boni2004} analysed high resolution data of 10 giant stars in the Sgr
dSph galaxy
and obtained abundances of O, Mg, Si, Ca and Fe. They concluded that a
substantial
metal rich population exists in the Sgr dSph.
High dispersion data of a total of 13 giant stars in Draco, Ursa Minor and
Sextans dSph galaxies were analysed by \citet{shet2001}. They found large
internal dispersions in metallicity of all three galaxies. They also found that
the relative abundances of $\alpha$--elements, [$\alpha$/Fe], are lower
in dSph galaxies compared with those found in the halo field stars over the
same range in metallicity, which hints a non-negligible contribution of
Type Ia Supernovae (SN Ia) during early stage of chemical enrichement,
but the number of stars observed is too small to conclude.
\citet{shet2003} and \citet{tols2003} carried out extensive abundance
analyses of
15 red giant stars in the Sculpter, Fornax, Carina, and Leo I dSph galaxies and
discussed the implications for understanding the history of galaxy formation.
\citet{shet2003} found that certain abundance patterns appear to be very
similar
between these four dSph galaxies and Ursa Minor, Draco, Sextans, and Sagittarius
dSph galaxies examined in the literature; i.e., iron---peak elements, 
second {\it s}--
and {\it r}--process elements all show Galactic halo-like patterns.
The $\alpha$--elements, however, can vary from galaxy to galaxy.
Sculpter, Leo I, Sextans, Ursa Minor, and Sagittarius dSph galaxies show a
slightly
decreasing [$\alpha$/Fe] pattern with increasing metallicity, while Fornax and
Draco show roughly constant [$\alpha$/Fe].
No uniform picture for nucelosynthesis in dSph galaxies yet appeared,
and clearly more abundance data are
desperately required.

In order to investigate in more detail the abundance patterns in giant
stars of northern dSph galaxies, we have initiated a program to observe
high resolution spectra of bright stars in the Ursa Minor dSph galaxy
in 2001. The previous study for this galaxy by \citet{shet2001} was based
on rather low S/N ($\sim 30$) spectra. We obtained
high resolusion spectra with higher quality (S/N$>50$) using the Subaru
Telescope to confirm the previous study and investigate the chemical
composition in more detail. In this paper, we report abundance analyses of
three stars observed in 2002.\\

\section {Observational Data}

Spectroscopic observations of three target stars (cos 4, cos 82, and cos 347;
designations are taken from \citet{cudw86})
and one reference star (M92-III-13, a member of the globular cluster M92)
  were carried out with the Subaru telescope
using the High Dispersion Spectrograph (HDS)  on 2002 May 16 and 17.
Figures 1 and 2 show our three target stars on the sky and on the
color-magnitude
diagram (CMD) of  the UMi  dSph galaxy.
These target stars are located near the tip of the red giant branch of
the UMi  dSph galaxy and show  somewhat peculiar color indices. Two
stars (cos 82 and cos 347) are slightly redder while cos 4 is slightly
bluer than the mean locus, suggesting that they might represent
higher or lower  metallicities with respect to the mean of the red
giants in the UMi  dSph galaxy.
  Data of two additional reference stars (BD +30$^{\circ}$2611 and HD 216143)
were obtained on 2001 June 3 using the same instrumental setup.
These reference stars were selected from a list of
well studied and relatively bright metal deficient giant stars \citep{bur2000}.
All of our reference stars are cooler than 4500 K in {\it $T_{\rm eff}$}, have
log {\it g} values smaller than 1.0, and their metallicity, [Fe/H], range from
-1.4 to -2.5.

The echelle grating
of this spectrograph is a mosaic of two 31.6 gr mm${}^{-1}$  gratings, and the
dispersion is 1 \AA~  mm${}^{-1}$  at 4300 \AA. The detector is a mosaic
of two 4k x 2k EEV CCD's with 13.5 $\mu$m pixels. We used a slit width of
1".0 (0.5 mm) and the 2x2 binning mode, which enabled us to achieve a
spectral resolution
of about 45000 by a 3.5 pixels sampling.
Technical details and the performance of the spectrograph are described
in \citet{nogu02}. Our observations covered the wavelength region from
4400 \AA~ to 7160 \AA~ with a gap between 5720 -- 5800 \AA.
Multiple exposures of 1800 sec were obtained for our target stars.  The journal
of our observation is given in table 1.
For flat-fielding of the CCD data, we obtained Halogen lamp exposures
(flat images) with the same setup as that for the object frames.

The reduction of two-dimensional echelle spectral data (bias subtraction,
flat-fielding, scattered-light subtraction, extraction of spectral data,
and wavelength calibration) was performed using the IRAF software package
in a standard manner. Spectral data extracted from multiple object images
were averaged in order to improve the signal-to-noise (S/N) ratio.
The wavelength calibration was done using the Th-Ar
comparison spectra obtained during the observations. The measured
FWHM of the weak Th lines is 0.13 \AA~ at 6000 \AA, and the resulting
resolution is around 46000. The S/N ratios of the resulting spectra
were measured at several continuum windows between 6100 \AA~ and 6200 \AA .
The averaged S/N ratio (per pixel) ranges from 50 to 60 for three UMi stars.
Those of the comparison stars are between 190 and 360 in the same wavelength
region.
Measurements of radial velocities of three UMi stars were carried out using
the D lines of Na~{\sc i}. All three UMi stars show large negative velocities
  (ranging from -235 to -255 km s${}^{-1}$), consistent with the mean value of the
  six UMi stars given in \citet{shet2001}. \\

In order to illustrate the quality of our data, a small section of the spectra
of three target stars together with three reference stars covering the region
between 6160 \AA~ and 6171 \AA~ is shown in figure 3. In this region,
we find five Ca~{\sc i} lines as well as a Na~{\sc i} line at 6160.75 \AA.
Additionally, we find a Pr~{\sc ii} line at 6165.89 \AA~ in cos 82 and cos 347,
which is not visible in the reference stars. We identify an
Er~{\sc ii} line at 6170.06 \AA~in cos 82.
On the other hand, we notice only one line
(Ca~{\sc i}   6162.18 \AA) in cos 4, which suggests that the star is
very metal deficient.

\section {Abundance Analysis}

\subsection{Line Identification and EW Measurement}

In order to prepare a list of absorption lines to be used in abundance
analyses, we first
registered all symmetric and clean lines observed in the spectrum of the
reference star BD +30$^{\circ}$2611 between 4600 and 7100 \AA. We do not
use the
spectral region below 4600 \AA, because
the SN ratio of the data of UMi stars become very low in the region. Next,
we tried to find unblended absorption lines consulting the line list of
\citet{kubell95}. In the process, we use the spectrum synthesis program
SPTOOL developed by Y. Takeda (private communication) which incorporates
the line list and can simulate any required spectral segment using an
appropriate model atmosphere for any combination of assumed abundances. As
a result, we prepared a list of about 700 clean absorption lines which
includes 27 chemical elements (from O to rare earths).
The list contains 260 Fe~{\sc i} lines together with many lines of other iron
peak elements such as Cr~{\sc i} and Ni~{\sc i}. Nine rare earth elements
(from La to Er) are
included in the list. Then, we examined the spectra of three UMi stars
comparing with the list
of BD +30$^{\circ}$2611 and selected lines to be measured in each star. In this
process, we noticed
that lines of heavy rare earth elements (Eu, Gd, Dy, and Er) are
extraordinarily strong in
cos 82 as illustrated in figure 3. \citet{shet2001} found a very large
    abundance of  Eu in this star
(star 199 in their table 4C), but no abundances were reported for Gd, Dy,
and Er.
Equivalent widths were measured  with the program SPTOOL using a Gaussian
fitting technique.
Equivalent widths measured in the present analysis for two stars (cos 347 and
HD 216143) are compared with published data in figure 4. For cos 347, our
results are compared
with those given in \citet{shet2001}. Results for HD 216143 are compared
with data given in \citet{john2002}. In both cases, we can find no
systematic trend or offset. The scatter in  the comparison of cos 347 is
very large compared to that in the case of HD 216143.  The difference
clearly demonstrates the effect of the S/N ratio on the equivalent width
measurements.

\subsection{Atmospheric Parameters and the Fe Abundance}

Atmospheric parameters ({\it $T_{\rm eff}$},
log {\it g},  $\xi$${}_{t}$, and [Fe/H]) for each program star
have been determined spectroscopically based on the equivalent widths
for a set of  selected   Fe\,{\footnotesize I} and  Fe\,{\footnotesize II}
lines.
We use in the present analyses those lines listed in the tables of
critically evaluated
log {\it gf} values given in \citet{lam1996}. Measured equivalent widths of
Fe\,{\footnotesize I} and  Fe\,{\footnotesize II} lines used in the
analyses are
given in table 2.  Interpolating  model atmospheres given in  \citet{kuru93},
we tried to find a solution for each star which satisfies  the following
three requirements simultaneously.\\
(1) the abundances derived from selected  Fe\,{\footnotesize I} lines
should not show any dependence on the lower excitation potential ($\chi$)
(excitation equilibrium), \\
(2) the averaged abundances derived from Fe\,{\footnotesize I} and
Fe\,{\footnotesize II} lines should be equal (ionization equilibrium),
and \\
(3) the abundance derived from Fe\,{\footnotesize I} lines should show
no dependence on the equivalent width (matching of the shape of the
curve-of-growth,
i.e., independence on the equivalent widths).\\

For each of our target and reference stars, we constructed diagrams such as
shown in figure 5 (for HD216143) and figure 6 (for cos 4) in order to
examine the relations between the Fe abundance and the observed equivalent
widths and the lower excitation potential ($\chi$).
We repeated  calculations changing the relevant parameters until we obtain
a satisfactory fulfillment of the above three requirements.
In the process, we used only those lines with equivalent widths smaller
than  200 m\AA~ for each star.  Resulting parameters for six stars are
summarized in table 3. Uncertainties in {\it $T_{\rm eff}$}, log {\it g},
and $\xi$${}_{t}$ are estimated from these diagrams, while that in
[Fe/H] is the rms scatter in the derived abundances.
We list previously  published data of  atmospheric parameters for five
stars in table 4. Comparing these two tables, we generally find good
agreements in the obtained Fe abundances.
\citet{shet2001} obtained parameters ({\it $T_{\rm eff}$},
log {\it g}, $\xi$${}_{t}$, and [Fe/H]) for the two common stars in the UMi
dSph galaxy (cos 82 and cos 347) using nearly the same method as in our
analysis. Their results are in agreements with ours within the expected
errors including the reference star M92-III-13. \\

We find that one of the UMi stars, cos 4, is very metal poor ([Fe/H]$=-2.7$).
This star is analysed for the first time in our analysis.
This is the most metal--poor star found in UMi to date.
  \citet{shet2001}  found similarly metal--deficient stars  in
Draco (star 119, [Fe/H]$=-2.97$) and Sextans (S49, [Fe/H]$=-2.85$) dSph
galaxies. There is no object with such a low metalliciy in Carina,
Sculptor, Fornax,
and Leo I dSph galaxies analysed by \citet{shet2003}. Thus, the UMi star
cos 4 is one
of the lowest metallicty objects found in dSph galaxies so far.

\subsection {Elemental Abundances}

Abundances of elements other than  Fe have been obtained  starting  from
the list of
unblended lines  prepared for BD +30$^{\circ}$2611. Only those lines for
which we can find reliable data of transition probabilities (log {\it gf}
values)
were  selected from this list  to be used in abundance analyses. Generally,
we prefer to use
log {\it gf}  values given in the home page of NIST Atomic Spectra Database
of the
National Institute of Standards and Technology \citep{nist}, or in the VALD
atomic
line database \citep{kupka}.  When we can find data in both sources, we use
data
given in the NIST database. For  Si and several heavy elements, we use new
log {\it gf}  values found in recent papers.
We use log {\it gf} values given by \citet{boda2003}   for Si\,{\footnotesize I},
\citet{lawl2001a} for La\,{\footnotesize II},  \citet{biem2002} for
Ce\,{\footnotesize II},  \citet{biem1989}  for Sm\,{\footnotesize II},
and \citet{lawl2001b}  for Eu\,{\footnotesize II}.

Measured equivalent widths of absorption lines other than Fe are listed
in table 5 together with adopted log {\it gf}  values and their sources.
Averaged abundances of 26 elements in our six target stars are summarized
in table 6.

\subsubsection {Light Elements}

For the light element Na, we use equivalent widths of two subordinate lines
(5682.63 \AA~
and 5688.21 \AA) and the D lines to obtain the abundances. The damping
constants of the
D lines are taken from the VALD atomic line database \citep{kupka}. In cos
4 and cos 82,
we could use only the D lines, because the subordinate lines were found to
be too weak.
The derived abundances of Na ([Na/Fe]) in our target stars are generally
negative
(under--abundant) except for the reference star M92-III-13.
In cos 82,  we find a significant underabundance of Na, [Na/Fe]$=-1.11$.
The abundances of Mg are obtained from only one line of Mg\,{\footnotesize
I} at 5528.41
\AA. The resulting abundance of Mg ([Mg/Fe]) ranges from +0.4 to +0.7 dex in
the reference stars,
while it ranges from +0.1 to +0.4 dex in three UMi stars.
Thus, we find a slight under--abundance of Mg in UMi stars when compared to
  our reference stars. We use two weak lines of Si\,{\footnotesize I} to
obtain the abundance of
Si. For these lines, we adopt recently published solar log {\it gf} values
given in
\citet{boda2003}. The resulting abundances of Si ([Si/Fe]) show solar values
except for
cos 4, in which we find an over--abundance, [Si/Fe]$=+0.66$. Considering
the weakness of the
Si\,{\footnotesize I} lines and the relatively poor S/N ratio for cos 4,
the apparent
over--abundance of Si in this star should be interpreted with caution.
For Ca, we measured equivalent widths of at least 12 Ca\,{\footnotesize I}
lines.
The resulting abundances of Ca ([Ca/Fe]) in UMi stars show no significant
differences from the
reference stars. \citet{shet2001} noted that all their sample stars
belonging to three
dSph galaxies show a statistically significant under--abundances
of Mg and Ca when compared with  Galactic
halo stars with the same metallicity. They obtained a mean value of
[$\alpha$/Fe] of +0.13 $\pm$ 0.04 for UMi dSph stars. They noted that
the mean value of [$\alpha$/Fe] for halo field stars over the same
metallicity range is +0.28 $\pm$ 0.02 dex.  We obtained from
our three UMi stars a mean value of  [$\alpha$/Fe] to be +0.16,
which is nearly coincident with the result of \citet{shet2001}

\subsubsection {Iron Peak Elements}

For  V, Cr, Co, and Ni, our results of abundances ([X/Fe]) in three UMi
stars do not show
significant differences from the three reference stars. The abundances of
Mn are determined
from equivalent widths of nine Mn\,{\footnotesize I} lines. We find that
the Mn\,
{\footnotesize I} lines are strikingly weak in cos 4. In figure 7, we
compare the strongest
line of  Mn\,{\footnotesize I} in cos 4, cos 347, and BD +30$^{\circ}$2611. The
Mn\,{\footnotesize I}
line at 5394.68 \AA~ is fairly strong in the latter two objects, while the
line is invisible in
cos 4. Assuming an upper limit of the equivalent width of the
Mn\,{\footnotesize I} line to
be 10 m\AA, we estimate the upper limit of
the abundance of Mn in cos 4 to be [Mn/Fe] $\leq -0.70$. The abundance of
Mn found
in cos 4 is lower than in any other dSph stars analysed by
\citet{shet2001}. We can find no
dSph object which shows such a low abundance of Mn in \citet{shet2003},
either.\\
The abundances of Cu are obtained from the Cu\,{\footnotesize I} line at
5105.54 \AA.
The  Cu\,{\footnotesize I} line in cos 4 is compared with those observed in
cos 347 and BD$^{\circ}$+30  2611 in figure 8. We find that the
Cu\,{\footnotesize I} line at
5105.54 \AA~ is invisible in cos 4. From an estimated upper limit of the
equivalent width (10 m\AA) of the
line, we obtain the upper limit of the Cu abundance in cos 4 to be [Cu/Fe]
$\leq -0.76$.
The upper limit of Cu found for cos 4 is among the lowest values found in
\citet{shet2001}.

\subsubsection {Heavy Elements}

    From the measured equivalent widths of two Y\,{\footnotesize II} lines,
we find a low abundance of Y in cos 4, [Y/Fe]$=-0.56$. The abundances of Y
found in cos 82 and
in cos 347 are in agreement with results given in  \citet{shet2001}.
We obtain a high abundance of Zr in cos 82 from the Zr\,{\footnotesize II}
line at
5112.28 \AA. No data of the Zr abundances in dSph stars are given in
\citet{shet2001}, nor
in \citet{shet2003}. The abundances of Zr ([Zr/Fe]) in three reference stars are
found to coincide
with the solar value.

We use three Ba\,{\footnotesize II} lines in deriving the abundances of Ba.
The Ba\,{\footnotesize II} line at 6496.90 \AA~ in three stars
(cos 4, cos 82 and cos 347) is compared with that in the reference star
BD$^{\circ}$+30 2611 in figure 9. We notice that the Ba\,{\footnotesize II}
line is very weak in cos 4. The
resulting abundance of Ba in cos 4 is [Ba/Fe]$=-1.28$, which is the lowest
abundance of Ba
in all dSph stars analysed by \citet{shet2001}.

Absorption lines of nine rare earth elements (La through Er) have been
surveyed in our
three dSph stars. We could identify and measure only two weak lines of
Nd\,{\footnotesize II} in
cos 4. On the other hand, we identified absorption lines
of heavy rare earths Gd (Z = 64), Dy (Z =66), and Er (Z =68) in cos 82.
Figure 10 shows the identification of the Dy\,{\footnotesize II}  line at
5090.39 \AA~ in
cos 82. Although the Dy\,{\footnotesize II} line is weakly visible in the
reference star
BD +30$^{\circ}$2611, the line is extraordinally stong in cos 82.
Each three Gd\,{\footnotesize II} and Er\,{\footnotesize II} lines are clearly
identified in cos 82, and abundances of these heavy rare earth elements
have been determined for the first time in this star.

For several odd--Z rare earth elements such as La, Pr and Eu, we have to
take the effects of hyper--fine splitting (hfs) into account in abundance
determinations. Data of hfs for La\,{\footnotesize II},
Pr\,{\footnotesize II}, and Eu\,{\footnotesize II} lines were provided by
\citet{lawl2001a}, \citet{aoki2001}, and \citet{lawl2001b},
respectively. For each line of these elements, we computed curves of
growth with and without the effect of hfs for each target star.
Then, the necessary correction factors have been evaluated and applied to
each line to obtain the final abundances. When a line has a large splitting
and large observed equivalent width (stronger than 120 m\AA),
a correction factor as large as $-0.9$ dex has to be applied.

\section {Discussion}

Two stars (cos 82 = 199; cos 347 = 297) are common in
\citet{shet2001} and our sample. Both analyses give very similar
abundances for iron and $\alpha$--elements and show no sign of
systematic difference, confirming basic results of \citet{shet2001},
although the signal--to--noise ratio are not exactly the same.
In figure 11, [Ca/Fe] and [Mn/Fe]
for three stars (cos 4, cos 82, cos 347) are plotted against [Fe/H],
together with four stars (177, K, O, 168) taken from \citet{shet2001}.
We find that [Ca/Fe] of UMi dSph stars are systematically lower
than Galactic metal-poor stars at [Fe/H] $< -1.5$. This could be explained
1) if SNe Ia contributed in much earlier stage of chemical enrichment
(due to lower SFR) and/or 2) if a later stage of small star
formation events had fewer high-mass Type II Supernovae (SNe II),
thus resulting in lower [Ca/Fe] values as suggested by \citet{shet2003}.
Taking the decline of [Ca/Fe] in Ursa Minor dSph galaxy
as a sign of SNe Ia explosions,
\citet{ikuta2002} claimed that the SFR should be 20 -- 40 times
lower than that in the solar neighbourhood, if the Salpeter initial
mass function is assumed.
With such a very low SFR, the star formation should
last at least 4 -- 6 Gyrs in order to explain the observed metallicities
in cos 82 ([Fe/H]$=-1.51$) and cos 347 ([Fe/H]$=-1.67$) derived in this study.

Mn is supposed to come from SNe Ia (e.g., \cite{naka1999}),
thus its abundance would provide a crucial test for distinguishing the
two cases given above. We  find that both cos 82 ([Mn/Fe]$=-0.15$)
and cos 347 ([Mn/Fe]$=+0.05$) show noticeable enhancement of Mn with
respect to the Galactic halo stars of similar metallicity, which may suggest
that SNe Ia had contributed already significantly at
this early stage of chemical enrichment.
We should point out that cos 4 is enriched in Mg ([Mg/Fe]$=0.31$),
Si ([Si/Fe]$=0.66$), and Ca ([Ca/Fe]$=0.21$), while showing
considerably low Mn abundance ([Mn/Fe] $<-0.70$),
implying that SNe Ia had little contribution at this very beginning
of the enrichement.

If the star formation has been inefficient as suggested by \citet{ikuta2002},
the chemical enrichement in Ursa Minor dSph, or in dSph galaxies
in general, might have been inhomegeneous and the observed stars may
reflect the local enrichement due to particular supernovae exploded
in the vicinity of these stars.
We have compared the abundance patterns of cos 4, cos 82,
and cos 347 from Mg to Zn with metal-poor stars ($-2<$[Fe/H]$<-1$;
\cite{grat89};  \cite{sneden91};  \cite{grasne88};  \cite{grasne91}; 
\cite{grasne94};
\cite{edva93}; \cite{mac1995}; \cite{nissen97};  \cite{stephens99})
and extremely metal-poor stars ($-4<$[Fe/H]$<-2$;
\cite{cayrel}) in the Milky Way. We may conclude that
there is no clear sign for the influence of local supernovae explosions
in the abundance patterns from Mg to Zn in these three stars:
1) [Na/Fe], [Mg/Fe], [Si/Fe], [Ca/Fe], and [Cr/Fe] of cos 4 are normal at 
[Fe/H]$=-2.7$.
On the other hand, [Ti/Fe] and [Mn/Fe] are definitely lower in cos 4 than in the
Galactic metal--poor giants. [Ni/Fe] and [Zn/Fe] in cos 4 are also
slightly lower than the Galactic mean values at [Fe/H]$=-2.7$.  2) cos 82 
and cos 347
show very similar abundance patterns to each other. With respect to the
metal-poor stars in the Milky Way, these stars have similar values of [Mg/Fe],
[Co/Fe], [Ni/Fe], [Cu/Fe], and [Zn/Fe], but significantly lower values of
[Si/Fe], [Ca/Fe], and [Ti/Fe]. [Cr/Fe] and [Mn/Fe] are enhanced and
are close to the disc stars of higher metallicity ([Fe/H]$>-1$).
The abundance patterns from Mg to Zn suggest that
SNe Ia had already contributed significantly to the enrichment before
cos 82 and cos 347 were formed.
We therefore tentatively conclude that the three stars we observed
were not affected by local supernovae explosions, instead they
all keep the record of global enrichment history of Ursa Minor dSph.

Figure 12 shows relative abundances of neutron capture elements (Y through Er)
in the heavy element-enhanced stars cos 82 ([Eu/Fe]$=0.97$) and cos 347
([Eu/Fe]$=0.61$). We compare the observed abundance patterns with the
solar--system {\it r}--process and {\it s}--process abundance patterns.
We take the total solar--system abundances from \citet{gre98} and use the 
{\it r}--process
and {\it s}-process fractions in the solar--system given by \citet{bur2000}.
Surprisingly, we find that the general features of heavy rare earth elements
agree quite well with the scaled solar system {\it r}--process abundance curve.
The agreement of the abundance pattern with that of
the solar system ${\it r}$--process component was already suggested by
Shetrone et al. (2001), but our study clearly confirms this including
the heavier elements Gd, Dy, and Er.
This implies that the heavy rare earth elements of these
stars are of {\it r}--process origin, and the contribution from the
{\it s}--process is still considerablly small, even though the
metallicities of these stars are remarkablly high.
The abundance patterns of heavy rare earth elements are very close to that of
Galactic metal-poor stars, in particular, we find the ratio [Ba/Eu]$=-0.59$
and $-0.43$ for cos 82 and cos 347, respectively, again showing that
the ratios are clearly associated with the {\it r}--process (as is shown in
\citet{mcw98}), rather than
the {\it s}--process, although the contribution from the {\it s}--process may
not be negligible in cos 347.

The abundance patterns of individual elements prevent us from deriving a fully
consistent picture of chemical enrichement in the Ursa Minor dSph galaxy:
low abundances of $\alpha$ -- elements and high abundance of Mn imply a
significant contribution from SNe Ia (lifetime of the progenitor
$\sim 10^9$ yrs) at low metallicity ([Fe/H]$< -1.5$),
suggesting rather long period of star formation with very low SFR.
On the other hand, the contribution from the {\it s}--process
(i.e., from AGB stars of lifetime $\sim 10^8$ yrs) is
still considerablly small even at [Fe/H]$=-1.5$, which seems to suggest
that the time scale of star formation in the Ursa Minor dSph galaxy is
very short. The latter possibility was proposed by \citet{tsujimoto}.
If one assumes that there are two distinct classes of massive supernovae
-- one produces and ejects {\it r}--process elements and the other does
not, and if one further assumes that SNe II with the main sequence mass of
$20-25$ M$_{\odot}$ are the dominant site for {\it r}--process nucleosynthesis
\citep{tsujimoto}, a sharp rise of [Ba/Fe] in dSph galaxies
should occur between [Fe/H]$\sim -2.5$ and  $\sim -2.0$,
which corresponds to massive explosions of these SNe II
whose progenitors' lifetimes are an order of a few $10^7$ years.
Unfortunately, very small number of observed stars prevents us
from further analysis. A systematic study of extremely metal-deficient stars
([Fe/H]$<-2.0$) with much higher signal-to-noise spectra
would certainly provide a critical understanding of the early enrichment
history of dSph galaxies.

\section {Summary}

Our main findings concerning the abundances of the three UMi dSph
stars are as follows. 

\begin{enumerate}

\item We find a very metal poor star cos 4 in which Fe
is more under--abundant than other stars in the UMi dSph analysed so far.

\item In cos 4, we find that three elements Mn, Cu and Ba are extra--ordinally
deficient. 
The underabundance of Mn implies that SNe Ia had little contribution in this star. 
Y appears to be under--abundant in this star, too.

\item We find that the light element Na is significantly under--abundant in
cos 82. 

\item Abundances of  $\alpha$ elements (especially Ca) in cos 82 and cos 347 
are found to be lower, while that of Mn in these two stars to
 be  higher than the corresponding values in Galactic metal poor stars.

\item At the same time, we find large excesses of heavy neutron-capture elements
in these two stars  
with the general abundance pattern similar to the scaled solar system
{\it r}--process abundance curve. 

\end{enumerate}

These results appear to be rather puzzling: low abundances of $\alpha$ 
-- elements
and high abundance of Mn seem to sugggest a significant contribution of SNe Ia
at [Fe/H]$=-1.5$, while there is no hint of {\it s}--process (i.e., AGB
stars) contribution even at this metallicity, probably suggesting a peculiar
nucleosynthesis history of the UMi dSph galaxy.

\vskip 15mm

We thank Dr. A. Schweizter for providing us with unpublished data of proper
motion and photometry of the UMi dSph and  also thank
all staff members of the Subaru telescope, NAOJ,
for their help during the observation. Thanks are also due to M. Ohkubo
for her help in preparing the manuscript.
This research was partly supported  by a
Grant-in-Aid, No. 13640230 (to N.A.) and No. 15540236 (to K.S.) from the
Ministry of Education, Culture, Sports, Science and Technology, Japan.

\onecolumn
\setcounter {table} {0}
\begin{table}[h]
  \caption{Observational Log.}\label{first}
  \begin{center}
\begin{tabular}{cccccc}
      \hline\hline
      Object   & $\it V$   & $\it B - V$  & Exposure  & Obs. Date & SN (per
      pixel) \\
      &         &             & (Min.)      &         & (at 6100
      \AA)  \\
      \hline
      HD 216143   &   7.82      &    0.94            &   4              & 2001,
      June, 3 &   350              \\
      BD
      +30$^{\circ}$2611   &  9.14      &    1.16            &  10              &
      2001,
      June, 3 &   360              \\
      M92-III-13       &   12.03      &  1.28            &  20          & 2002,
      May, 16&   190               \\
      cos 347         &  16.94      &    1.52            &   150            &
      2002, May, 16 &   55              \\
      cos 4            &   16.73      &    1.33            &   210           &
      2002, May, 16 &   50              \\
      cos 82          &    17.2      &    1.36            &   180             &
      2002, May, 16, 17 &   60          \\
      \hline
    \end{tabular}
  \end{center}
\end{table}
\setcounter {table} {1}
\begin{table}[h]
      \caption{Equivalent widths of Fe~{\sc i} and Fe~{\sc ii}
lines.}\label{second}
\tiny
     \begin{center}
        \begin{tabular}{ccrrrrrrrr}
\hline\hline
      ion  & Wevelength & $\chi$    & log {\it gf} &        cos 4&   cos 82 &
cos 347
            &  BD +30$^{\circ}$2611  & HD 216143     & M92-III-13    \\
      & ($\rm\AA$)      & (eV) & & & & & &  &\\
\hline
Fe~{\sc i}      &       4966.087        &       3.332   &       -0.84   &
40      &               &               &               &       69      &
82      \\
            &       4969.916        &       4.217   &       -0.75   &
&               &               &       69      &       19      &       28
\\
            &       4973.101        &       3.960   &       -0.85   &
&               &               &               &       27      &       37
\\
            &       5002.789        &       3.396   &       -1.44   &
&               &               &               &       27      &       40
\\
            &       5014.941        &       3.943   &       -0.27   &       30
&       121     &               &       106     &       52      &       58
\\
            &       5022.236        &       3.984   &       -0.49   &
&       91      &               &               &       40      &       47
\\
\hline
        \end{tabular}
      \end{center}
\end{table}
\newpage
\setcounter {table} {1}
\begin{table}[h]
      \caption{continued.}
\tiny
     \begin{center}
        \begin{tabular}{ccrrrrrrrr}
\hline\hline
      ion  & Wevelength & $\chi$    & log {\it gf} &        cos 4&   cos 82 &
cos 347
            &  BD +30$^{\circ}$2611  & HD 216143     & M92-III-13    \\
      & ($\rm\AA$)      & (eV) & & & & & &  &\\
\hline
Fe~{\sc i}      &       5618.631        &       4.209   &       -1.26   &
&               &               &       38      &       7       &
\\
            &       5633.975        &       4.991   &       -0.12   &
&               &               &       43      &       8       &
\\
            &       5679.025        &       4.652   &       -0.77   &
&               &               &       39      &       7       &       11
\\
            &       5686.524        &       4.548   &       -0.45   &
&               &               &       59      &       12      &       18
\\
            &       5883.813        &       3.960   &       -1.21   &
&       60      &       65      &       65      &       14      &
\\
            &       5916.249        &       2.453   &       -2.99   &
&       78      &       96      &               &       20      &       44
\\
\hline
        \end{tabular}
      \end{center}
\end{table}
\setcounter {table} {2}
\begin{table}[h]
      \caption{Atmospheric parameters.}\label{third}

      \begin{center}
      \begin{tabular}{ccccc}
\hline\hline
      Object   & {\it $T_{\rm eff}$}   & log {\it g} & $\xi$${}_{t}$ & [Fe/H]
\\
               &     (K)               &             & (km s${}^{-1}$)
&                \\
\hline
HD 216143   &   4450 $\pm$ 100  & 0.8 $\pm$ 0.2 &  1.8 $\pm$ 0.2 & -2.18
$\pm$ 0.11  \\
BD +30$^{\circ}$2611 &   4250 $\pm$ 100  &  0.7 $\pm$ 0.2 &  2.0 $\pm$ 0.2
& -1.52
$\pm$ 0.10  \\
M92-III-13  &   4200 $\pm$ 100  &  0.5 $\pm$ 0.2 &  1.9 $\pm$ 0.2 & -2.15
$\pm$ 0.13  \\
cos 347     &   4050 $\pm$ 130  &  0.3 $\pm$ 0.3 &  2.1 $\pm$ 0.3 & -1.63
$\pm$ 0.19  \\
cos 4       &   4300 $\pm$ 130  &  0.3 $\pm$ 0.3 &  2.0 $\pm$ 0.3 & -2.66
$\pm$ 0.18  \\
cos 82      &   4300 $\pm$ 110  &  0.3 $\pm$ 0.3 &  2.0 $\pm$ 0.3 & -1.53
$\pm$ 0.18  \\
\hline
        \end{tabular}
      \end{center}
\end{table}

\setcounter {table} {3}
\begin{table}[h]
      \caption{Previously published atmospheric parameters.}\label{fourth}

      \begin{center}
      \begin{tabular}{cccccc}
\hline\hline
      Object   & {\it $T_{\rm eff}$}   & log {\it g} & $\xi$${}_{t}$ & [Fe/H]
& Reference*   \\
               &     (K)               &             & (km s${}^{-1}$)
&      &          \\
\hline
HD 216143   &   4582            &  1.46          &  2.5           &  -2.24
& 1      \\
                &   4525            &  0.80          &  1.8           &
-2.18      & 2      \\
                &   4400            &  0.70          &  1.80          &
-2.26      & 3      \\
                &   4525            &  1.0           &  2.85          &  -2.1
        & 4      \\
               &   4525            &  0.80          &  1.80          &  -2.00
       & 5      \\
               &   4500            &  1.0           &  1.60          &  -2.11
       & 6      \\
               &   4500            &  0.70          &  2.10          &  -2.10
       & 7      \\
BD +30$^{\circ}$2611
&   4275            &  0.80          &  2.1           &  -1.49
&  2     \\
                 &  4500            &  1.4           &  2.10          &
-1.2       &  4      \\
                 &  4275            &  0.80          &  2.10          &
-1.40      &  5      \\
                 &  4300           &  0.70          &  1.60          &
-1.41       & 6      \\
M92-III-13  &   4125           &  0.75          &  1.98          & -2.17
& 8      \\
                &   4180           &  0.10          &  2.15          &
-2.24       &  9     \\
                &   4125           &  0.75          &  2.00          &
-2.39       &  10    \\
                &   4175           &  0.20          &  2.10          &
-2.28       &  11    \\
cos 347     &   4075           &  0.40          &  2.30          & -1.68
&  11    \\
cos 82      &   4325           &  0.30          &  1.95          & -1.45
&  11    \\
\hline
        \end{tabular}
      \end{center}
* References: 1. \citet{fran1996},  2. \citet{pila1996}, 3.
\citet{shet1996},    4. \citet{ful2000},
5. \citet{bur2000}, 6. \citet{mis2001}, 7. \citet{john2002}, 8.
\citet{carr1997},
                          9. \citet{shet1998}, 10 \citet{sned00}, 11.
\citet{shet2001}
\end{table}

\newpage
\setcounter {table} {4}
\begin{table}[h]
      \caption{Equivalent widths of other elements.}\label{fifth}
\footnotesize
     \begin{center}
        \begin{tabular}{llccccccccc}
\hline\hline
      ion  & Wevelength & $\chi$    & log {\it gf} &  Ref        &cos 4&   cos
82 &   cos 347
            &  BD +30$^{\circ}$2611  & HD 216143     & M92-III-13    \\
      & ($\rm\AA$)      & (eV) & & & & & &  & &  \\
\hline
Na~{\sc i}      &       5682.633        &       2.102   &       -0.700  &
1       &               &               &       43      &       31      &
8       &       28      \\
            &       5688.205        &       2.104   &       -0.457  &       1
&               &               &       59      &       50      &       12
&       47      \\
            &       5889.951        &       0.000   &       0.112   &       1
&       186     &       243     &       347     &       293     &       222
&               \\
            &       5895.924        &       0.000   &       -0.191  &       1
&       166     &       233     &       332     &       269     &       196
&               \\
Mg~{\sc i}      &       5528.405        &       4.346   &       -0.620  &
2       &       64      &       150     &       174     &       182     &
116     &       137     \\
Si~{\sc i}      &       5948.541        &       5.082   &       -1.110  &
3       &       14      &       32      &       41      &       52      &
18      &               \\
\hline
        \end{tabular}
      \end{center}
\end{table}
\newpage
\setcounter {table} {4}
\begin{table}[h]
      \caption{continued.}
\footnotesize
     \begin{center}
        \begin{tabular}{llccccccccc}
\hline\hline
      ion  & Wevelength & $\chi$    & log {\it gf} &  Ref        &cos 4&   cos
82 &   cos 347
            &  BD +30$^{\circ}$2611  & HD 216143     & M92-III-13    \\
      & ($\rm\AA$)      & (eV) & & & & & &  & &  \\
\hline

            &       5192.969        &       0.021   &       -1.005  &       1
&       60      &       227     &       247     &       192     &       88
&       133     \\
            &       5210.386        &       0.048   &       -0.883  &       1
&       67      &       151     &       239     &       177     &       83
&       138     \\
            &       5219.700        &       0.021   &       -2.292  &       1
&               &       62      &       143     &       96      &       13
&       45      \\
            &       5426.237        &       0.021   &       -3.006  &       1
&               &       44      &       78      &       52      &       4
&       20      \\
            &       5460.469        &       0.048   &       -2.804  &       1
&               &       42      &       91      &       59      &       6
&       25      \\
            &       5471.197        &       1.443   &       -1.389  &       1
&               &       14      &       21      &       23      &
&               \\
\hline
        \end{tabular}
      \end{center}
\end{table}
\newpage
\setcounter {table}{4}
\begin{table}[h]
      \caption{continued.}
\footnotesize
     \begin{center}
        \begin{tabular}{llccccccccc}
\hline\hline
      ion  & Wevelength & $\chi$    & log {\it gf} &  Ref        &cos 4&   cos
82 &   cos 347
            &  BD +30$^{\circ}$2611  & HD 216143     & M92-III-13    \\
      & ($\rm\AA$)      & (eV) & & & & & &  & &  \\
\hline

            &       5432.546        &       0.000   &       -3.795  &       1
&               &               &       162     &       122     &       13
&       39      \\
            &       5470.637        &       2.164   &       -1.702  &       2
&               &       38      &       57      &       57      &       6
&       10      \\
            &       5516.774        &       2.178   &       -1.847  &       1
&               &       29      &       48      &       36      &       4
&       8       \\
            &       5537.760        &       2.187   &       -2.017  &       2
&               &       40      &       51      &       36      &       4
&       8       \\
            &       6013.513        &       3.072   &       -0.252  &       1
&               &       56      &       70      &       59      &       10
&       9       \\
            &       6016.673        &       3.073   &       -0.216  &       2
&               &       65      &       72      &       68      &       13
&       15      \\
\hline
        \end{tabular}
      \end{center}
\end{table}
\newpage
\setcounter {table}{4}
\begin{table}[h]
      \caption{continued.}
\footnotesize
     \begin{center}
        \begin{tabular}{llccccccccc}
\hline\hline
      ion  & Wevelength & $\chi$    & log {\it gf} &  Ref        &cos 4&   cos
82 &   cos 347
            &  BD +30$^{\circ}$2611  & HD 216143     & M92-III-13    \\
      & ($\rm\AA$)      & (eV) & & & & & &  & &  \\
\hline
Pr~{\sc ii}     &       5219.045        &       0.795   &       -0.235  &
1       &               &       80      &               &       23      &
5       &       3       \\
            &       5259.740        &       0.630   &       0.082   &       1
&               &       112     &       46      &       27      &       6
&       6       \\
            &       5292.619        &       0.648   &       -0.294  &       1
&               &       83      &       44      &       24      &       4
&       3       \\
            &       5322.772        &       0.483   &       -0.461  &       2
&               &       119     &       49      &       32      &       7
&       8       \\
            &       5352.398        &       0.483   &       -0.727  &       2
&               &       66      &       16      &       13      &
&               \\
Nd~{\sc ii}     &       4943.899        &       0.205   &       -1.514  &
2       &               &       64      &               &       26      &
5       &       4       \\
\hline
        \end{tabular}
      \end{center}
References: 1. \citet{nist}         2.  \citet{kupka}      3.
\citet{boda2003}   4. \citet{lawl2001a}
     5. \citet{biem2002}      6. \citet{biem1989}       7. \citet{lawl2001b}  \\
\end{table}
\newpage
\setcounter {table} {5}
\begin{table}[h]
      \caption{Abundances in Target Stars.}\label{sixth}
\footnotesize
     \begin{center}
        \begin{tabular}{lccccccccc}
\hline\hline
           &\multicolumn{3}{c}{cos 4} & \multicolumn{3}{c}{cos 82} &
\multicolumn{3}{c}{cos 347}                          \\
            &               &       S.D.    &       N       &               &
S.D.         &       N       &       & S.D.  &       N      \\
\hline
$[$Fe~{\sc i}/H$]$      &       -2.65   &       0.15    &       61      &
-1.51   &       0.22    &       63      &       -1.67   &       0.14    &
58      \\
$[$Fe~{\sc ii}/H$]$     &       -2.70   &       0.17    &       6       &
-1.55   &       0.17    &       10      &       -1.53   &       0.16    &
9       \\
$[$Na/Fe$]$     &       -0.25   &               &       2       &
-1.11   &               &       2       &       -0.27   & 0.17 &        4
\\
$[$Mg/Fe$]$     &       0.31    &               &       1       &
0.14    &               &       1       &       0.39    &               &
1       \\
$[$Si/Fe$]$     &       0.66    &               &       2       &
-0.06   &               &       2       &       0.17    &               &
2       \\
$[$Ca/Fe$]$  &  0.21    &       0.24    &       12      &       0.04    &
0.19    &       16      &       0.19    &       0.16    &       17      \\
\hline
        \end{tabular}
      \end{center}
\end{table}
\newpage
\setcounter {table} {5}
\begin{table}[h]
      \caption{continued.}
\footnotesize
     \begin{center}
        \begin{tabular}{lccccccccc}
\hline\hline
           &\multicolumn{3}{c}{BD +30$^{\circ}$2611} & \multicolumn{3}{c}{HD
216143} &
\multicolumn{3}{c}{M92-III-13}                                    \\
            &               &       S.D.    &       N       &               &
S.D.         &       N       &       & S.D.  &       N      \\
\hline
$[$Fe~{\sc i}/H$]$      &       -1.52   &       0.09    &       74      &
-2.12   &       0.08    &       98      &       -2.18   &       0.12    &
76     \\
$[$Fe~{\sc ii}/H$]$     &       -1.51   &       0.08    &       11      &
-2.16   &       0.07    &       15      &       -2.18   &       0.12    &
11     \\
$[$Na/Fe$]$     &       -0.31   &               &       2       &
-0.15   &       0.03    &       4       &       0.42    &               &
2      \\
$[$Mg/Fe$]$     &       0.43    &               &       1       &
0.64    &               &       1       &       0.74    &               &
1      \\
$[$Si/Fe$]$     &       0.13    &               &       2       &
0.24    &               &       2       &       0.36    &               &
1      \\
$[$Ca/Fe$]$     &       0.21    &       0.14    &       17      &
0.26    &       0.09    &       16      &       0.29    &       0.10    &
16     \\
\hline
        \end{tabular}
      \end{center}
\end{table}

\onecolumn
\begin{figure}
      \begin{center}
        \FigureFile(170mm,10mm){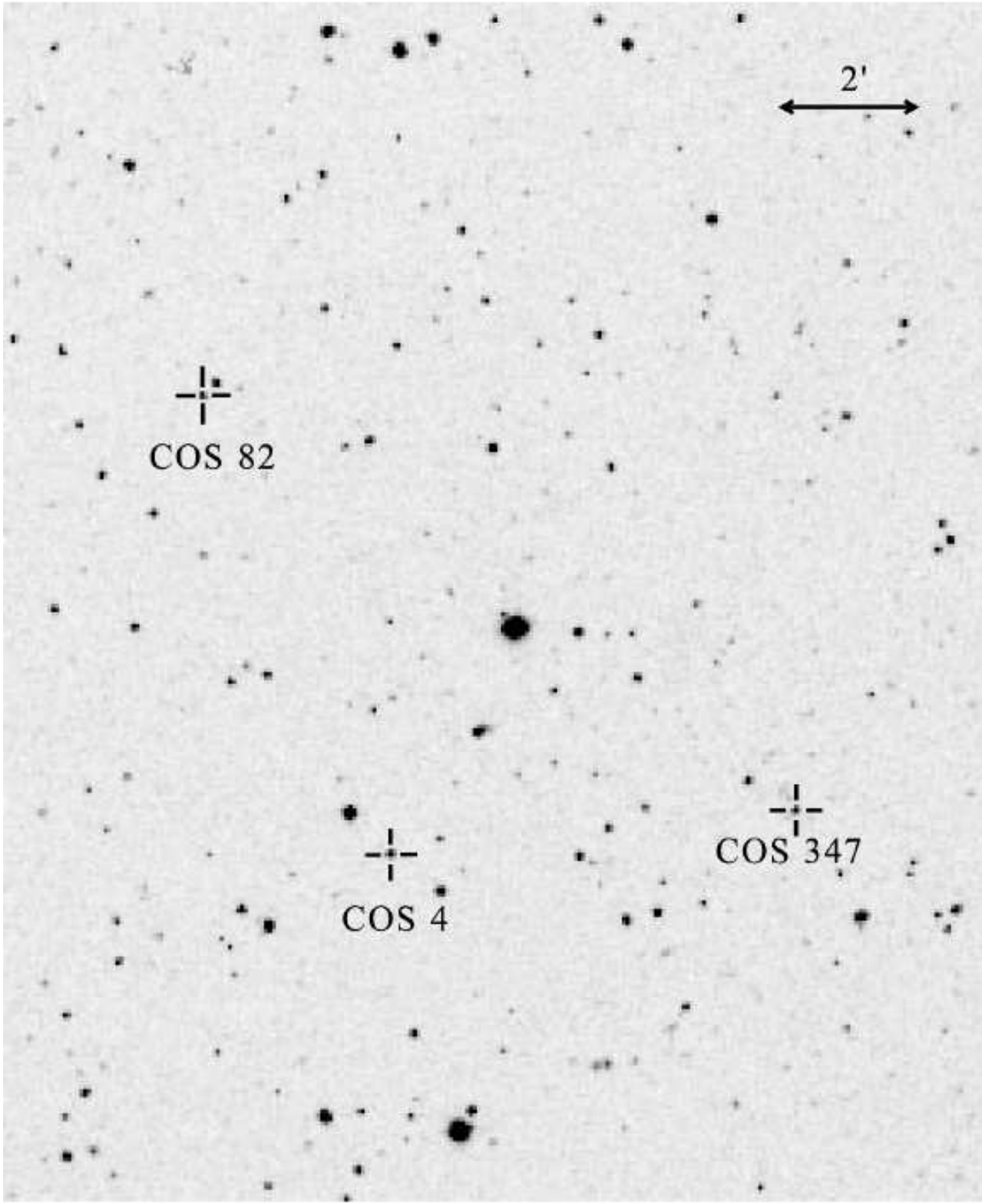}
      \end{center}
      \caption{A map of the sky area containing three target stars cos4,
         cos82, and cos 347. The figure is centered at R.A. (2000)
        \timeform{15h09m10s.0} and Dec (2000) \timeform{+67D12'52"}.}
     \end{figure}
\begin{figure}
      \begin{center}
        \FigureFile(170mm,10mm){fig2.epsi}
      \end{center}
      \caption{{\it B - V} color-magnitude diagram for the UMi dSph.
          Only stars with the membership probability higher than 90 \% are
          plotted on the figure. Diamond marks represent the RGB stars for which
          we took spectra. Photometric data and membership probability are taken
          from a catalogue prepared by A. Schweitzer (private communication). }
     \end{figure}
\newpage
\begin{figure}
      \begin{center}
        \FigureFile(170mm,10mm){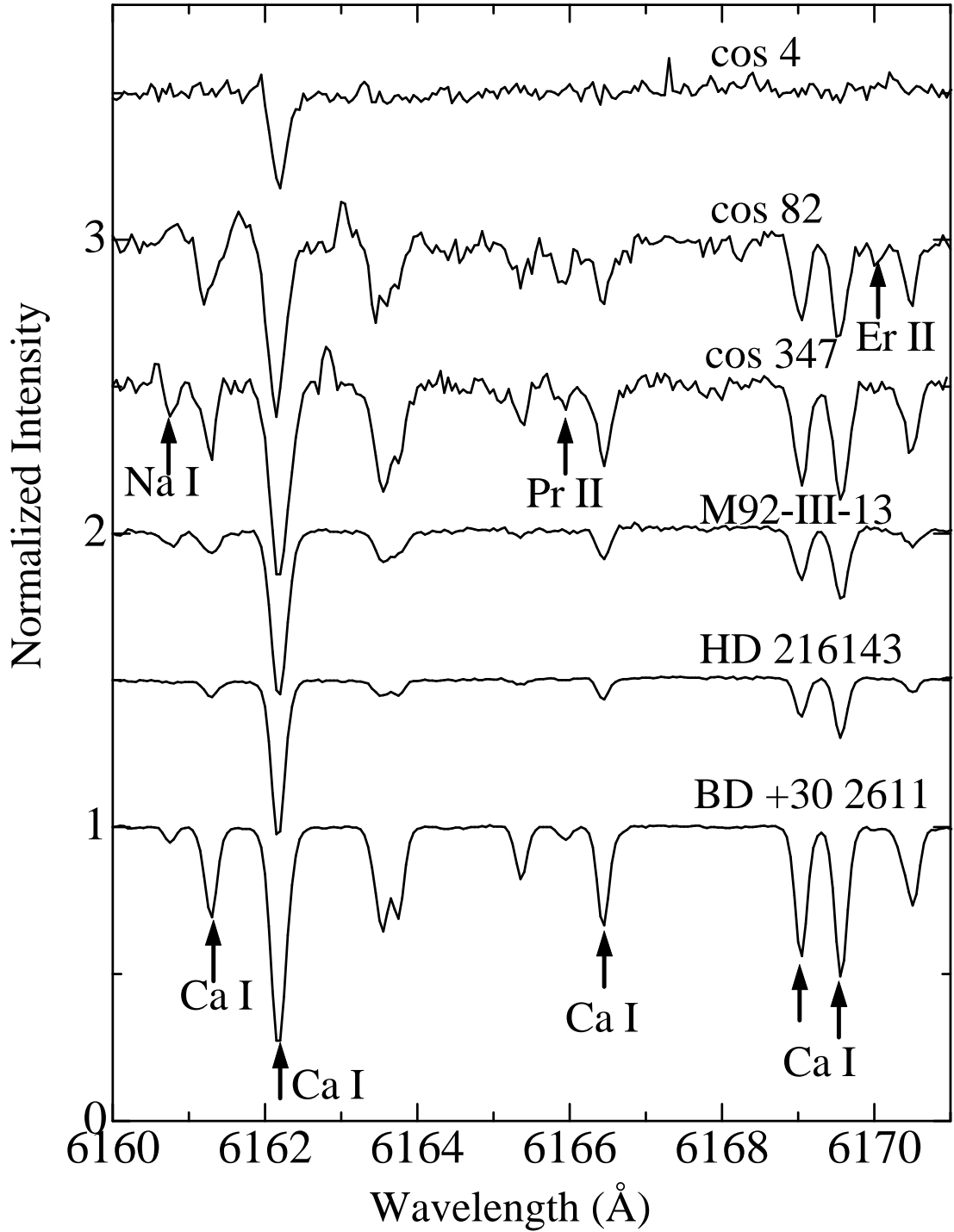}
      \end{center}
      \caption{A small portion of the observed spectra of target stars. Strong
lines of Ca~{\sc i}
       are identified for  the reference star BD +30$^{\circ}$2611.
Unusually strong
lines of heavy rare earth
         elements Pr and Er are identified in cos 82. In cos 4, only one line
(Ca~{\sc i} 6162 \AA) can be detected and it  suggests that
the star is very metal deficient.}
     \end{figure}
\begin{figure}
      \begin{center}
        \FigureFile(170mm,10mm){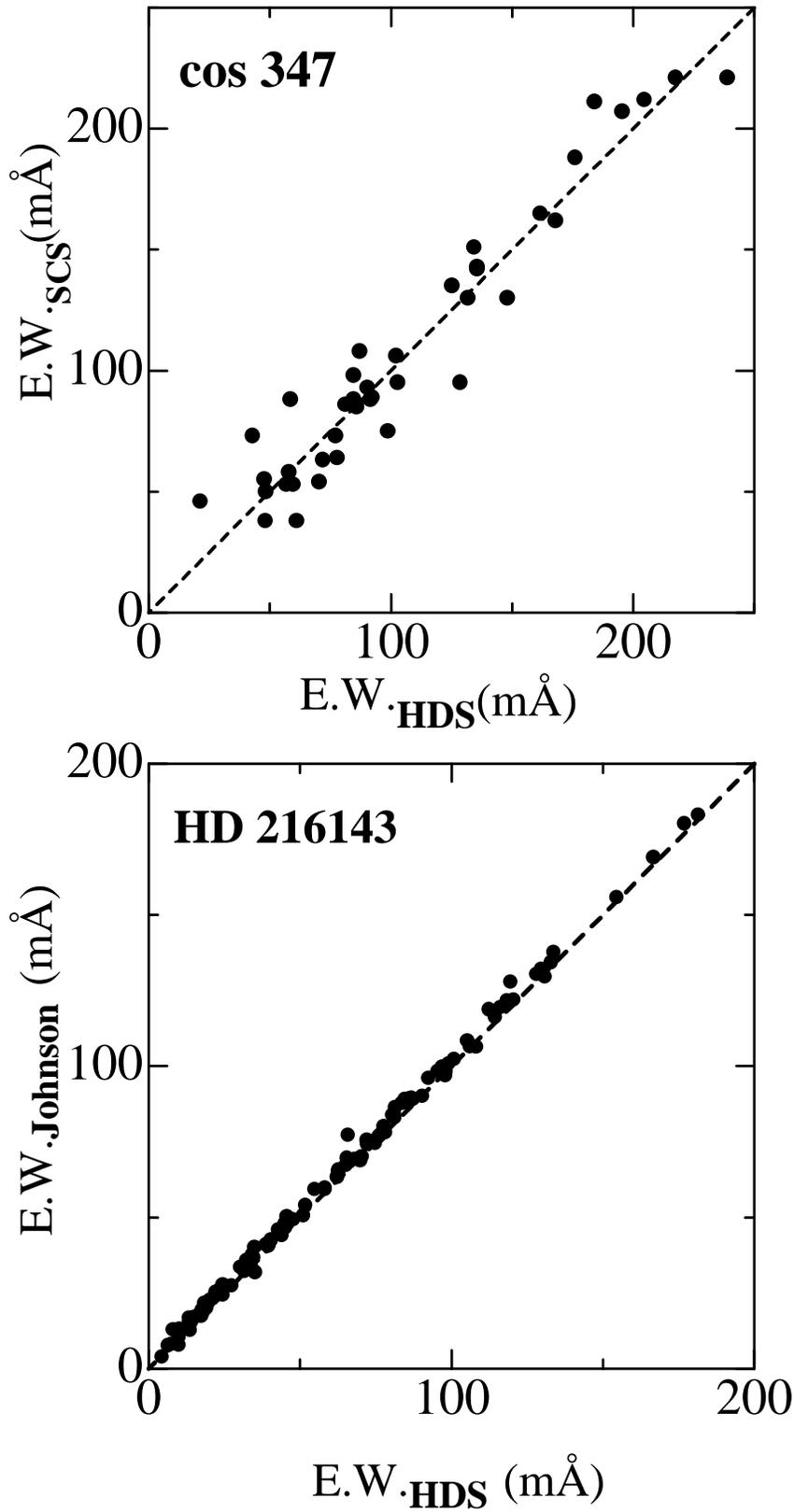}
      \end{center}
      \caption{Comparisons of observed equivalent widths. Our measurements
are compared
        with results given in \citet{shet2001} for cos 347 (upper panel) and
those given in \citet{john2002}
       for HD 216143 (lower panel).}
\end{figure}
\begin{figure}
      \begin{center}
        \FigureFile(170mm,10mm){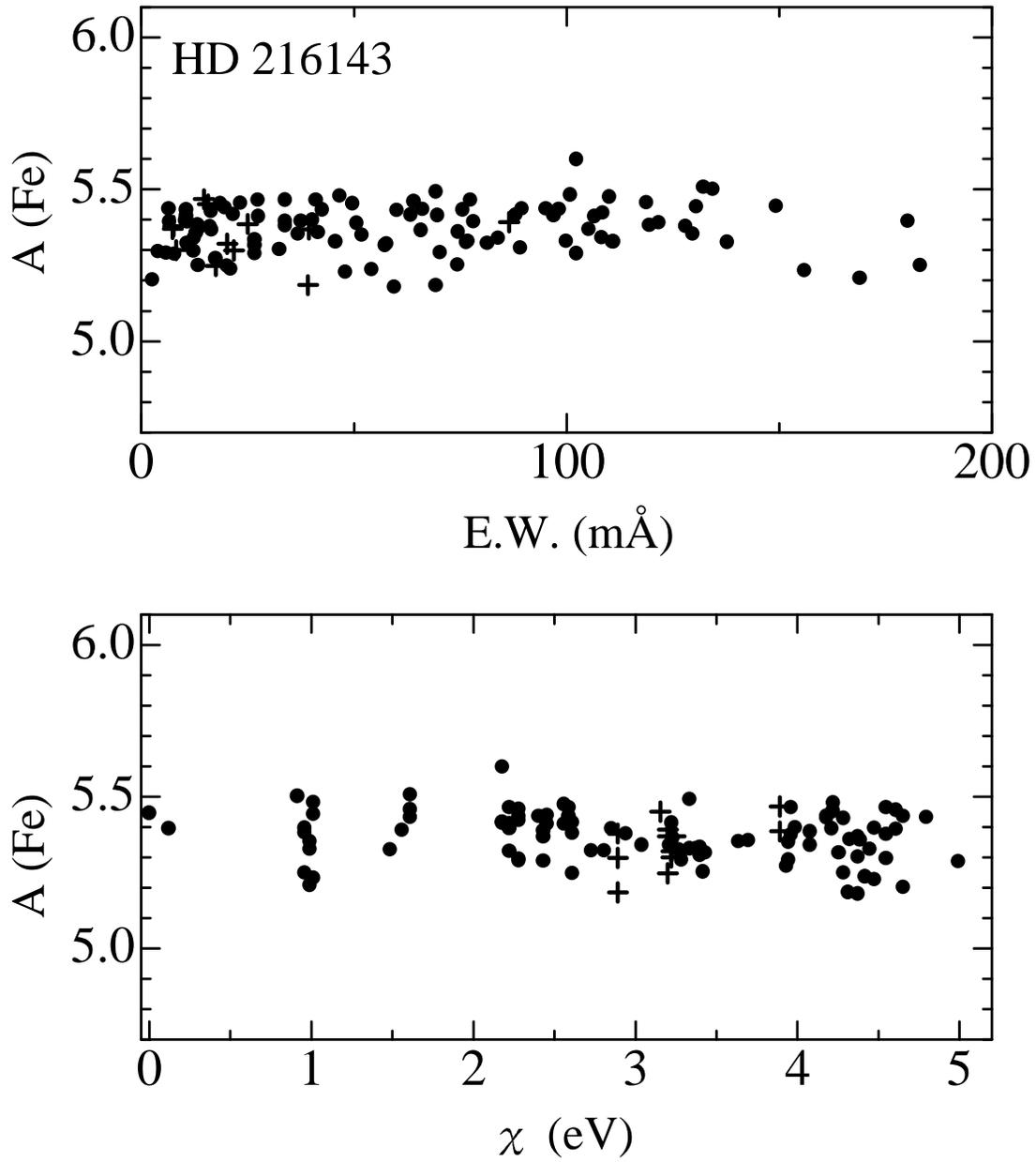}
      \end{center}
      \caption{Analyses of Fe~{\sc i} and Fe~{\sc ii} lines in HD 216143. The
relations between
        $\it  A$(Fe) and the equivalent
                widths (upper panel) and those with
                       the lower excitation potential  (lower panel) are
displayed.
               The Fe~{\sc i}  and  Fe~{\sc ii} lines are
               denoted by  filled circles  and plus signs, respectively.}

\end{figure}
\begin{figure}
      \begin{center}
        \FigureFile(170mm,10mm){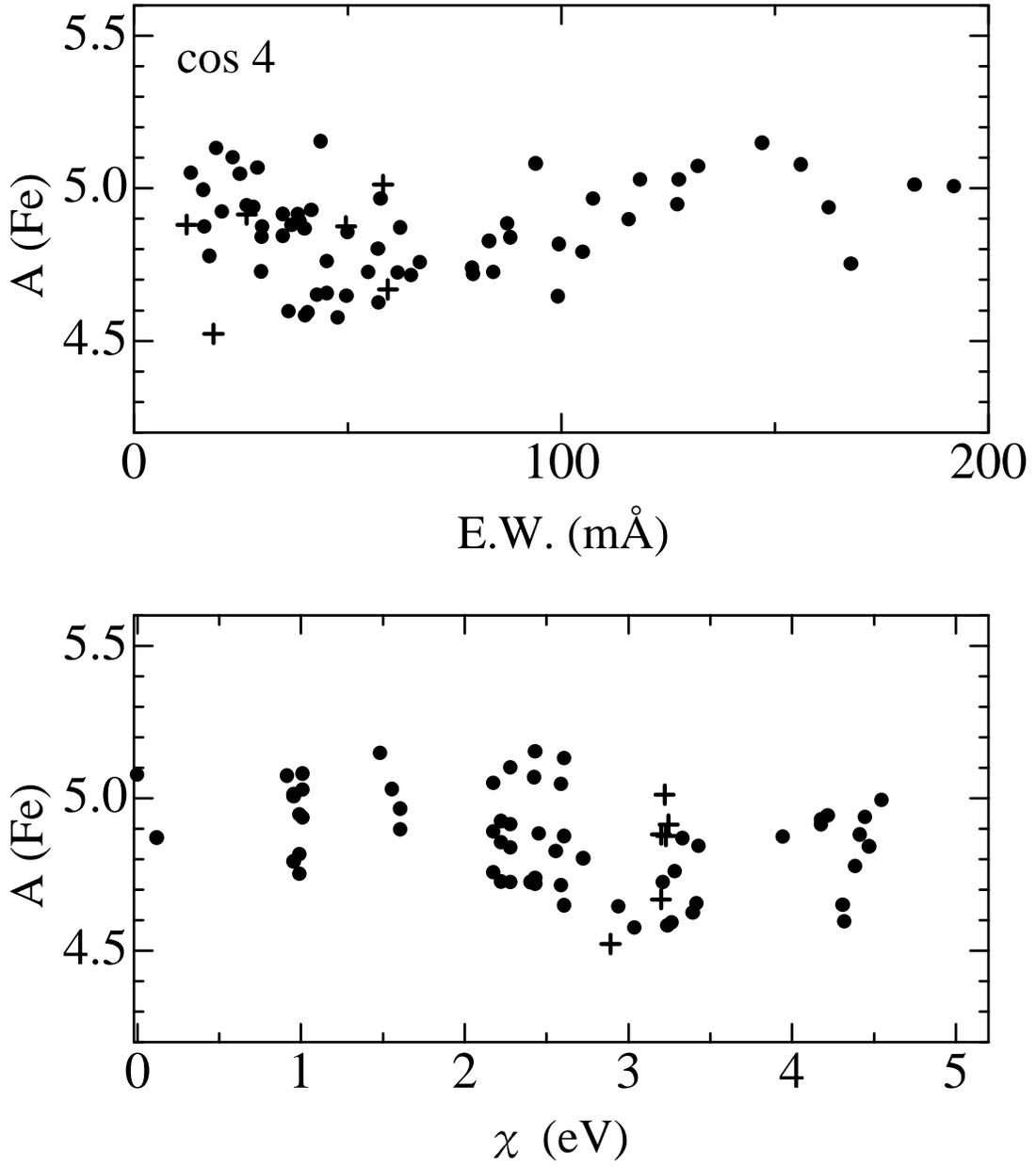}
      \end{center}
      \caption{Same as figure 3, but for cos 4}
\end{figure}
\begin{figure}
      \begin{center}
        \FigureFile(170mm,10mm){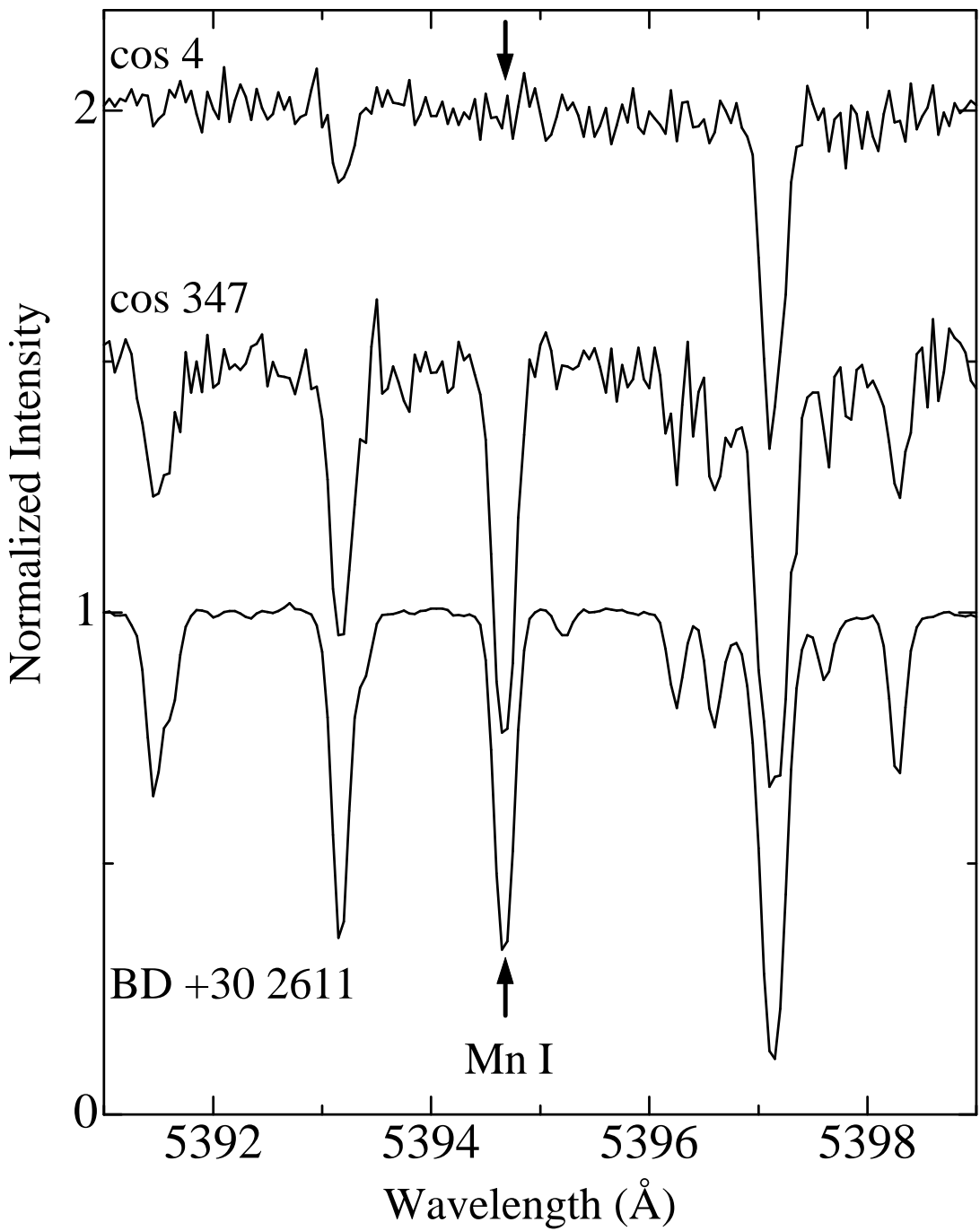}
      \end{center}
      \caption{The Mn~{\sc i} line at 5394.68 \AA. Note the weakness of the
line in
        cos 4.}
\end{figure}
\begin{figure}
      \begin{center}
        \FigureFile(170mm,10mm){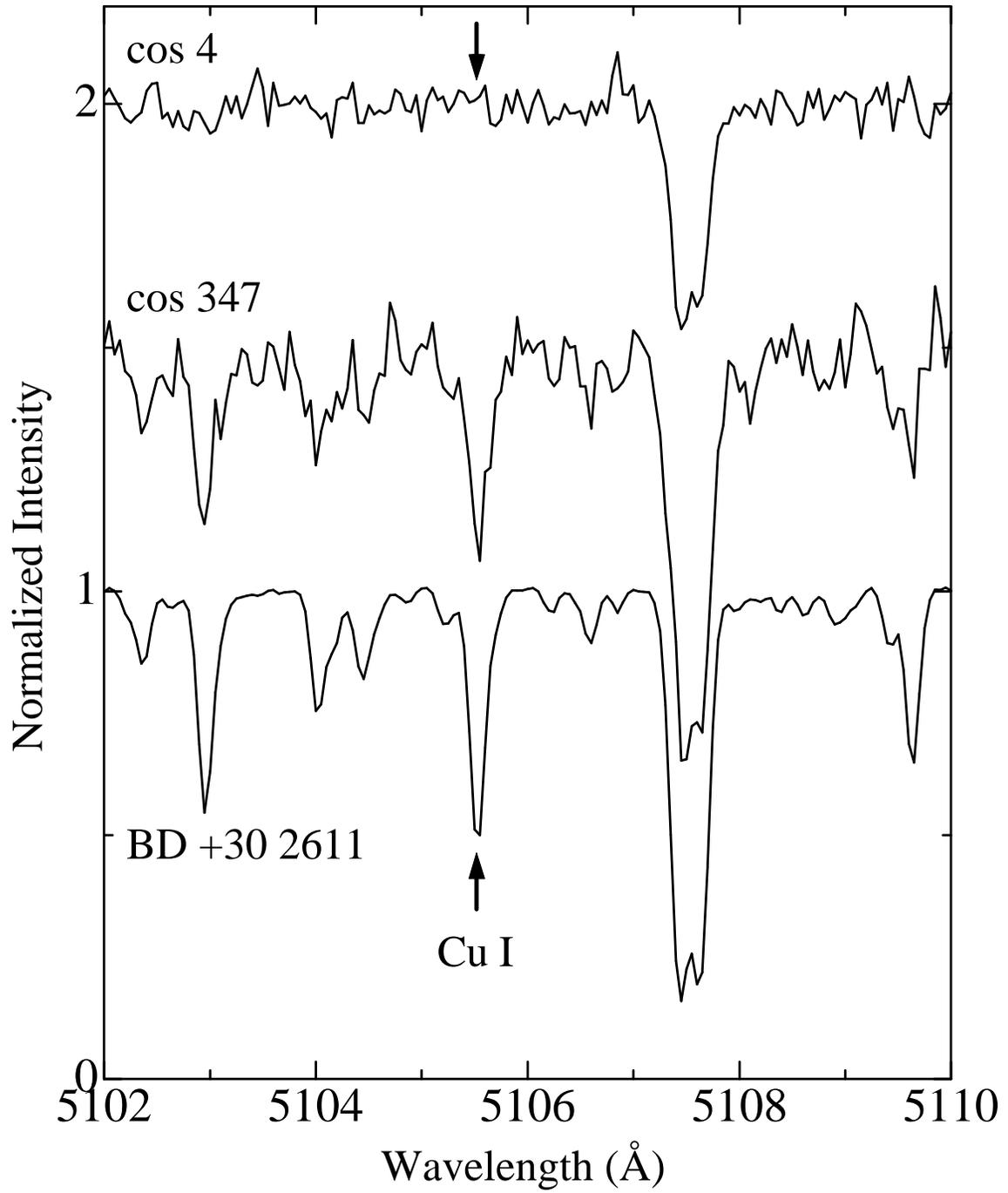}
      \end{center}
      \caption{The Cu~{\sc i} line at 5105.54 \AA. Note the weakness of the
line in
        cos 4.}
\end{figure}
\begin{figure}
      \begin{center}
        \FigureFile(170mm,10mm){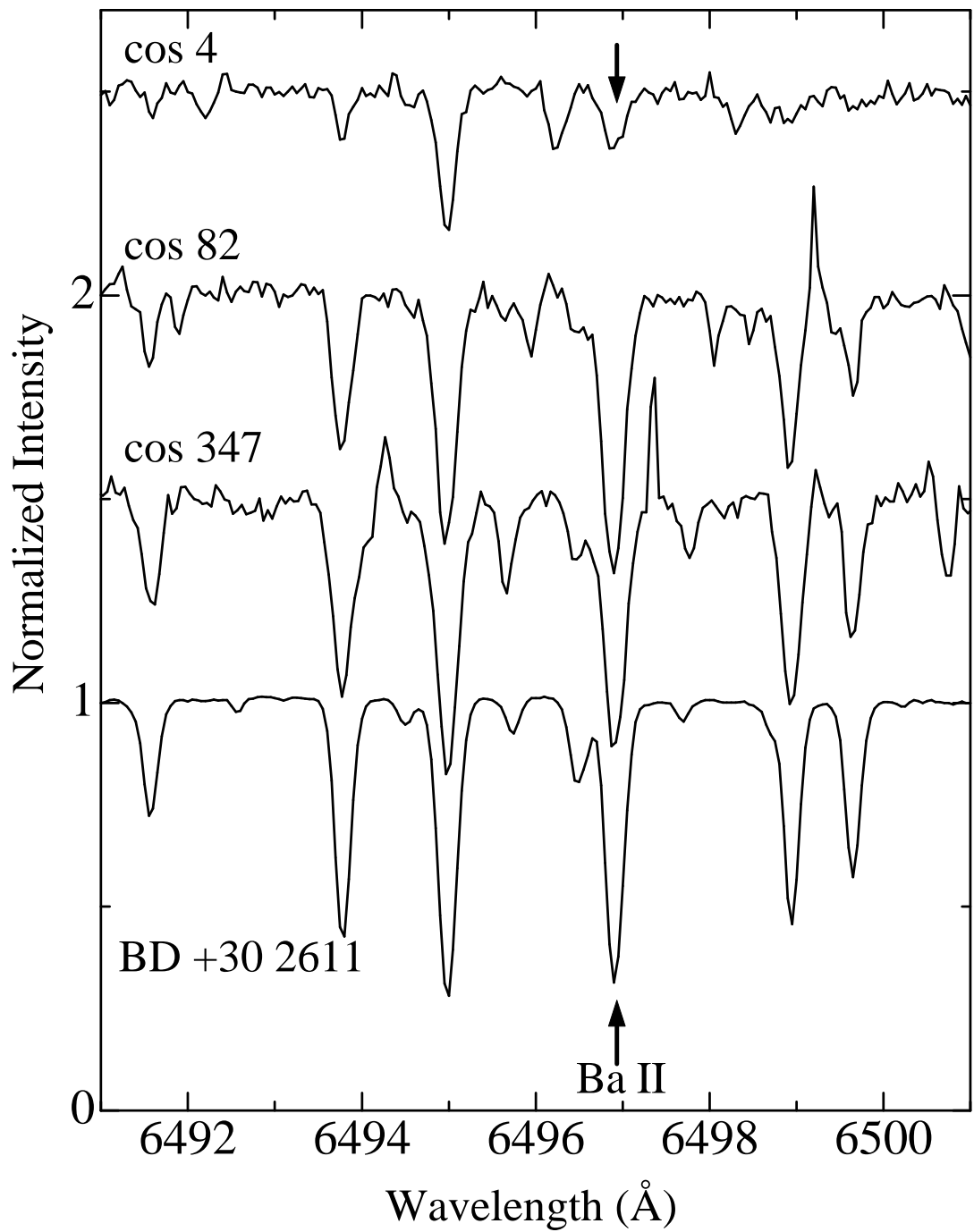}
      \end{center}
      \caption{The Ba~{\sc ii} line at 6496.90 \AA. Note the weakness of the
line in
        cos 4.}
\end{figure}
\begin{figure}
      \begin{center}
        \FigureFile(170mm,10mm){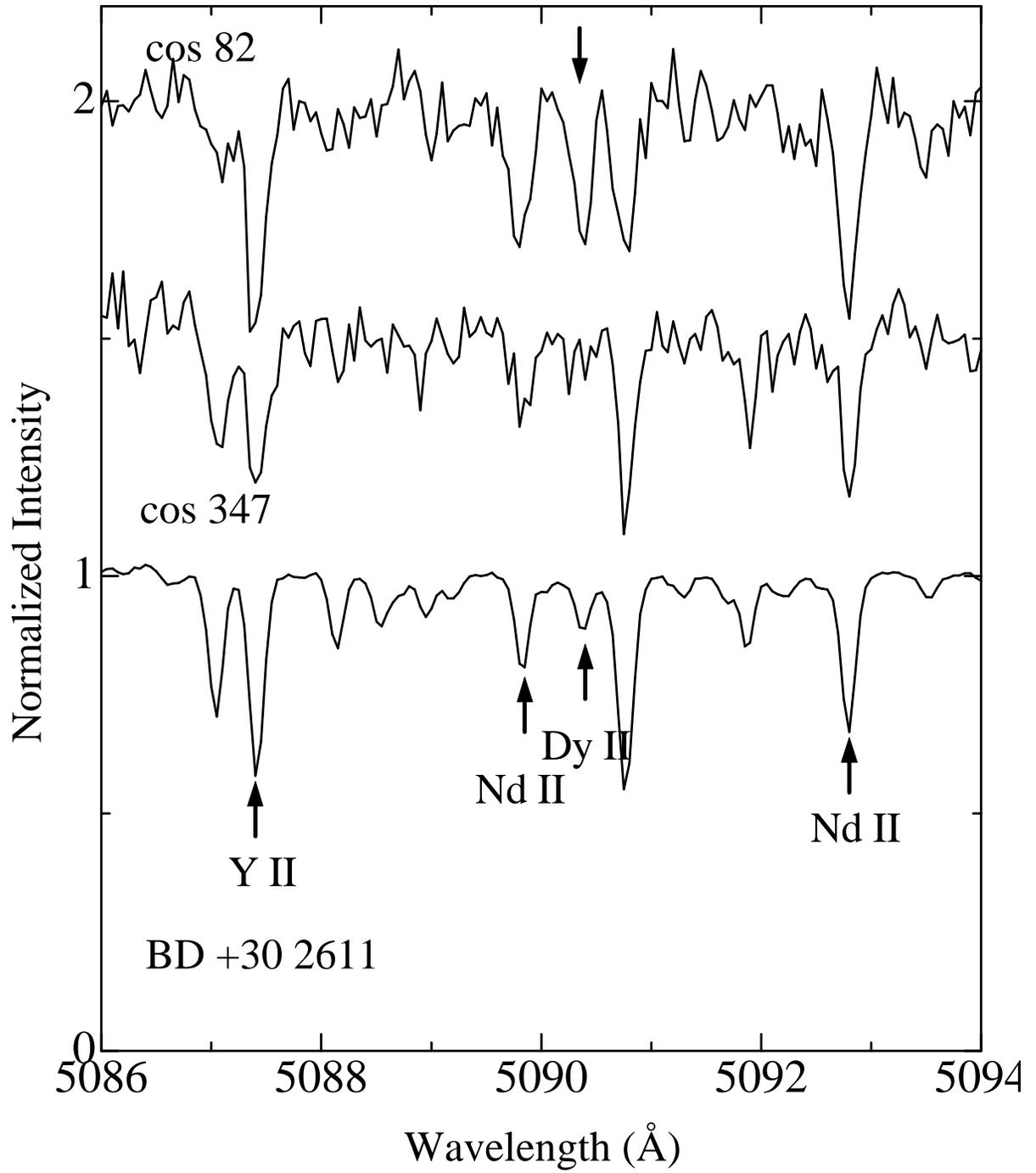}
      \end{center}
      \caption{The Dy~{\sc ii} line at 5090.39 \AA. The line is
extraordinally strong in cos 82.}
\end{figure}
\begin{figure}
      \begin{center}
        \FigureFile(170mm,10mm){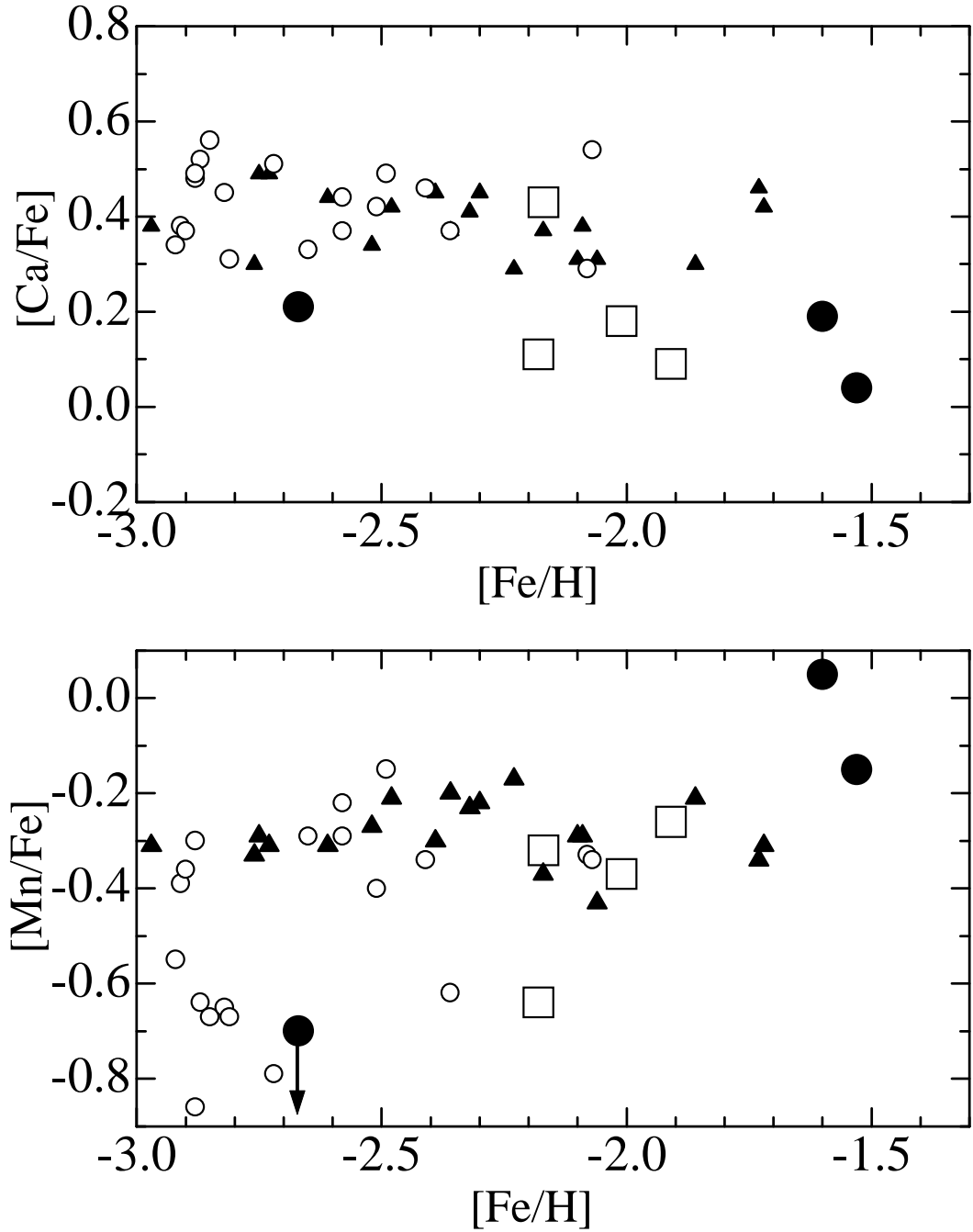}
      \end{center}
      \caption{[Ca/Fe] vs [Fe/H] (upper panel) and [Mn/Fe] vs [Fe/H] (lower
panel) relations
           for seven UMi  stars and for galactic metal poor stars. Three UMi
stars (cos 4, cos 82, and
           cos 347) are shown by large filled circles. Other UMi stars
analyzed in   \citet{shet2001}
           are shown by large open squares. Small filled triangles and small
open circles are galactic
           metal poor stars analyzed in \citet{john2002} and in
\citet{mac1995}, respectively.}
\end{figure}
\begin{figure}
      \begin{center}
        \FigureFile(170mm,10mm){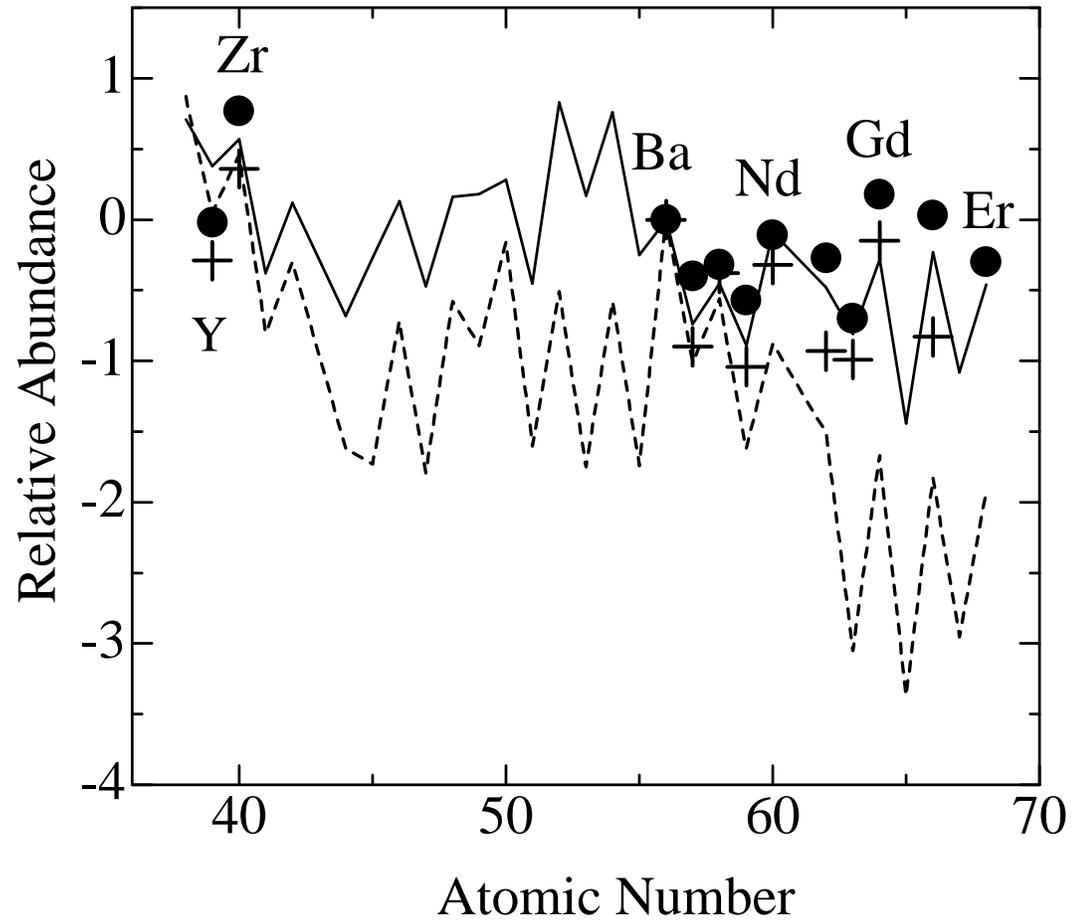}
      \end{center}
      \caption{Abundances of neutron capture elements (Y through Er) in cos 82
  (filled circles) and in cos 347 (plus signs).
         They are compared to the scaled solar system {\it r}  and {\it s} --process
         abundance curves, which are represented by solid and broken lines, respectively.
         Abundances are  normalized at Ba.}
\end{figure}

\end{document}